# HIGH POWER OPERATIONS OF LEDA*

L. M. Young, L. J. Rybarcyk, J. D. Schneider, M. E. Schulze[a], and H. V. Smith, Los Alamos National Laboratory, Los Alamos, NM 87544, USA

*Abstract*

The LEDA RFQ, a 350-MHz continuous wave (CW) radio-frequency quadrupole (RFQ), successfully accelerated a 100-mA CW proton beam from 75 keV to 6.7 MeV. We have accumulated 111 hr of beam on time with at least 90 mA of CW output beam current. The 8-m-long RFQ accelerates a dc, 75–keV, ~106-mA $H^+$ beam from the LEDA injector with ~94% transmission. When operating the RFQ at the RF power level for which it was designed, the peak electrical field on the vane tips is 33 MV/m. However, to maintain the high transmission quoted above with the CW beam, it was necessary to operate the RFQ with field levels ~10% higher than design. The RFQ dissipates 1.5 MW of RF power when operating with this field. Three klystrons provide the 2.2 MW of RF power required by the RFQ to accelerate the 100-mA beam. The beam power is 670 kW. Some of the challenges that were met in accelerating a 100-mA CW proton beam to 6.7 MeV, will be discussed.

## 1 INTRODUCTION

The LEDA RFQ [1,2] (see Figure 1) is the highest energy operational RFQ in the world [3-8]. Some of the unique features implemented in this RFQ to meet this goal include:

- It is over 9 wavelengths long, by far the longest 4-vane RFQ in the world.
- The transverse focusing at the RFQ entrance was reduced for easier beam injection.
- An electron trap is placed between the final focusing solenoid and the RFQ. The electron trap prevents the electrons in the beam plasma from flowing into the RFQ. With the electron trap turned off electrons flowing into the RFQ reduced the measured current as much as 25% from the correct value.
- The aperture and the gap voltage in the acceleration section are larger than in previous RFQ designs.
- The transverse focusing at the RFQ exit is reduced to match the transverse focusing strength in the coupled-cavity drift-tube linac [9].
- It is the first RFQ to utilize resonant coupling [10,11]. The RFQ is composed of four 2-m-long RFQs resonantly coupled together. The RF fields throughout its 8-m length are nearly as stable as the fields in the 2-m-long RFQs from which it is composed.
- RF power from 3 klystrons is coupled to the RFQ through 6 waveguide irises.

To implement the reduced focusing strength at the entrance of the RFQ and have adequate focusing in the interior of the RFQ, the transverse focusing parameter is increased smoothly from 3.1 to 7.0 over the first 32 cm of the RFQ. The focusing parameter is proportional to $V/r_0^2$ where $V$ is the voltage between adjacent vane tips and $r_0$ is the average aperture. The voltage is held constant in this region and the aperture is reduced to increase the focusing parameter. On entry, the beam is not yet bunched, allowing the use of weak transverse focusing. By the time the beam starts to bunch, the focusing is strong enough to confine the bunched beam. The reduced focusing strength at the entrance means the matched beam size is larger than it would have been without the reduced focusing strength. This allows the final focusing solenoid to be placed farther away from the RFQ. Without this feature proper placement of the final focusing solenoid is right at the RFQ entrance. With the focusing solenoid 30-cm from the RFQ, both simulations and experimental evidence indicate the beam becomes un-neutralized in the last 10 cm before the RFQ match point. Moving the final focusing solenoid 15-cm from the RFQ counteracts the effect of the defocusing from the beam's space charge.

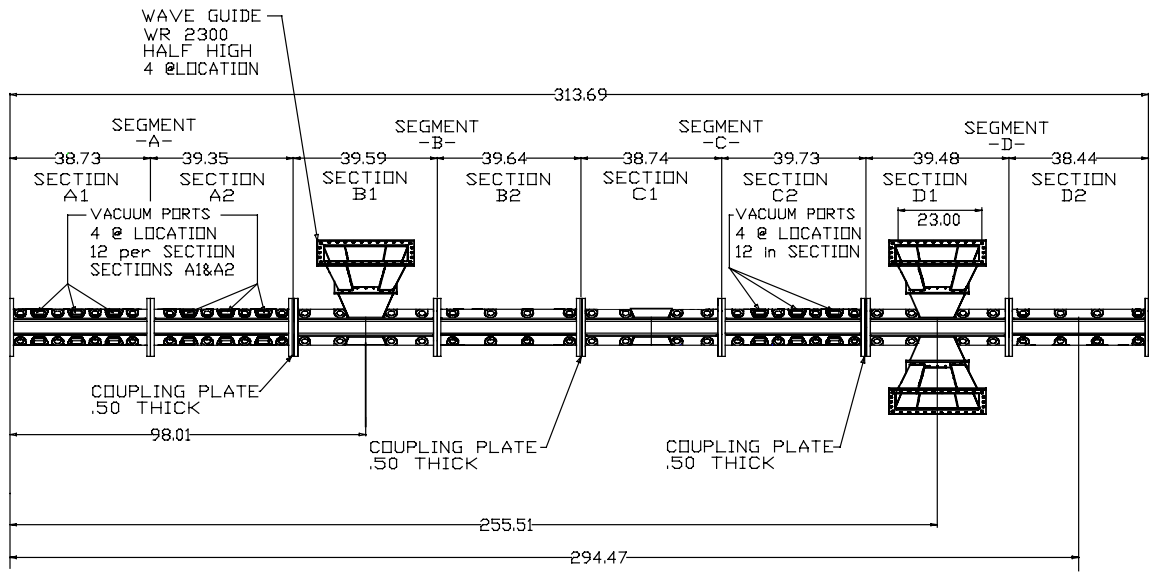

Figure 1: Line drawing of 8-m-long RFQ. This drawing shows the six RF-waveguide feeds used to power the RFQ; two on Section B1 and four on section D1.

*Work supported by the US Department of Energy.

[a] General Atomics, Los Alamos, NM 87544 USA.

## 2 RFQ DESIGN

### 2.1 Acceleration Section

In a typical RFQ that has constant focusing strength and constant gap voltage, as vane modulation increases to accelerate the beam, the aperture shrinks and beam can be lost on the vane tips. In an RFQ, as the energy rises the cell length increases and, for a given modulation, the accelerating gradient decreases inversely with cell length. Since the maximum practical modulation is about 2, the RFQ would become very long if the gap voltage remained constant. To reduce beam loss and shorten the RFQ, we maintain a large aperture, and increase the vane voltage.

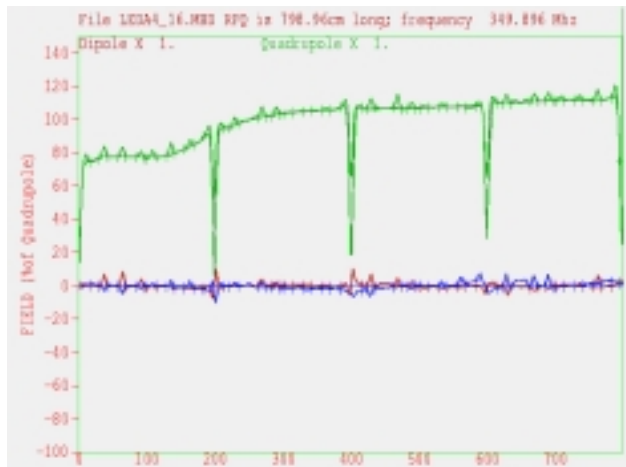

Figure 2: Bead perturbation measurement of the RF fields in the RFQ. The Quadrupole fields are normalized to 100%. The two residual dipole modes mixed with the RFQ fields are typically less then 2%.

The increased gap voltage substantially increases the accelerating field, thus shortening the RFQ. However, even with this increased gap voltage, eight meters of length is required to accelerate the beam to 6.7 MeV.

### 2.2 Resonate Coupling

A conventional 8-m-long, 350-MHz RFQ would not be stable. Small perturbations would distort the field distribution intolerably [10,11]. Therefore, four 2-m-long RFQs (labeled as segments A, B, C and D in Figure 1) are resonantly coupled to form the 8-m-long LEDA RFQ. The resonant coupling is implemented by separating the four 2-m RFQs by coupling plates. An axial hole in the coupling plate allows the vane tips to nearly touch. The capacitance between the vane tips of one RFQ and the next provides the RF coupling between the 2-m-long RFQ segments. The gap between the vane tips at the coupling joint is 0.32 cm. To minimize the effect of this gap on the beam, the gap is placed so that as a bunched beam pulse passes the gap, the RF electric field crosses zero. The RF field is in phase in all four segments. The "coupling mode" has a strong electric field across the 0.32-cm gap and has one longitudinal node in each 2-m RFQ segment. The coupling mode's longitudinal component of the electric field transmits RF power, and it is this mode which provides the stability to the fields. When the coupling mode is strongly excited (by a perturbation for example), a saw-tooth pattern can appear on the field distribution [11].

### 2.3 RFQ Fields

Figure 2 shows a measurement of the fields in the RFQ. The fields were measured with the bead perturbation technique in the magnetic field region close to the outer wall. In this measurement a bead is mounted on a plastic tape that is supported at the ends of the RFQ and at the coupling plates. The plastic tape with the bead moves on a pulley system and travels through all 4 quadrants of the RFQ. The bead perturbs the frequency of the RFQ in proportion to the stored energy of the magnetic field displaced by the bead. The frequency perturbation is measured versus bead position and the relative magnetic field strength is derived. In Figure 2, the bumps in the field are caused by local perturbations in the magnetic field near the slug tuners. A total of 128 tuners are used to "tune" the RFQ to the correct field distribution and frequency. The larger dips in the quadrupole magnetic field that occur every 200-cm are caused by the coupling plates. These dips and bumps do not appear in the electric field on axis. The RFQ is tuned using a "least squares" fitting procedure that minimizes the difference between the measured fields and the design fields. The slug tuner insertions are the parameters in this "least squares" fit.

The minimum aperture occurs about 1.4 meters into the RFQ, at the end of the gentle buncher. This is also the location where the transverse current limit goes through a minimum. Typically, the end of the gentle buncher is the RFQ choke point that determines the maximum current that can be accelerated (~200 mA for this RFQ) [7]. The theoretical current limit assumes ideal quadrupole fields and can only be used as a rough guide of the actual current limit.

## 2.4 RFQ Design

The RFQ was designed with the code PARMTEQM [12]. PARMTEQM is an acronym for "Phase and Radial Motion in a Transverse Electric Quadrupole; Multipoles". This code includes the effect of higher-order multipoles in the RFQ fields that are important in accurately predicting beam loss. In addition, PARMTEQM requires a realistic description of the input beam to accurately simulate beam losses in the RFQ when the input beam is not ideal. Simulations of the beam transport through the LEBT [3] with PARMELA [13,14] produces a more realistic distribution of particles for input into the RFQ simulation code than the ideal input distributions generated internally by PARMTEQM. Simulations of the RFQ with PARMTEQM predict output beam emittances in the range from 0.16 to 0.22 mm-mrad depending on the input distribution. The simulations also predict the x and y emittances to be the same. The measured x and y emittance [15,16] are 0.25 and 0.31 mm-mrad respectively. There are also minor differences between the predicted and measured Twiss parameters [15].

## 3 LEBT MODIFICATIONS

Until we added an electron trap described below to the LEBT, our transmission measurements were inaccurate. The input current was less than the current out of the RFQ. Electrons (from the beam plasma) flowing into the RFQ reduced the positive proton current measured by the toroid at the RFQ entrance. These electrons are responsible for neutralizing the proton beam space charge. We used the computer code PARMELA to perform a simulation of the beam traveling through the LEBT with 98% space charge neutralization, except for the last 10 cm in front of the RFQ. There, we made the simple assumption that the space charge neutralization changed linearly from 98% to 0 in 10 cm. The results of this simulation showed that the beam could not be properly "matched" into the RFQ. Space charge caused the beam to defocus so much in the last 10 cm that it no longer converged as it entered the RFQ. This limited the maximum beam current out of the RFQ to only 89 mA, equal to the maximum pulsed current that we obtained by August 23, 1999. Simulations showed that if we installed an electron trap to prevent the electrons from flowing into the RFQ and decreased the solenoid-to-RFQ distance from 30 to 15 cm, then the beam could be matched properly into the RFQ. The electron trap is a ring placed at the entrance of the RFQ. The potential on this ring, −1 kV, prevents low-energy plasma electrons from going through it, but does not affect the 75-keV protons. We made these changes to the LEBT on August 28–29, 1999.

## 4 TRANSMISSION THROUGH RFQ

Using the calculated beam from the modified LEBT, PARMTEQM predicts 93% transmission with the RFQ operating at the design field levels. This transmission is slightly less then the 95% transmission previously predicted with assumption that the space charge is 96% neutralized all the way to the RFQ match point [17]. RFQTRAK [18], a code that calculates the 3D space and image charge effects in an RFQ, agreed very well with PARMTEQM. The measured transmission has been as high as 94% at 100 mA when the RFQ fields are 10% above the design field strength. Figure 3 shows a comparison between the calculated and measured transmission as a function of field strength. This figure also shows an anomalous drop in transmission at the end of a 300-µs long pulse when the RFQ RF field strength is at or below the design field strength.

### 4.1 Ion Trapping in RFQ

Figure 4 shows the transmission in a 300-µs-long beam pulse when the RFQ fields are near the design field level.

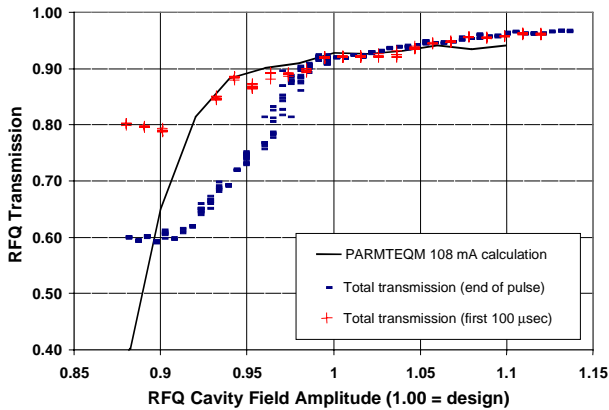

Figure 3: RFQ transmission versus cavity field using a 300-µs-long-beam pulse. The anomalous transmission drop occurs at slightly higher fields for longer pulses and CW beams. The calculated transmission is for accelerated beam only.

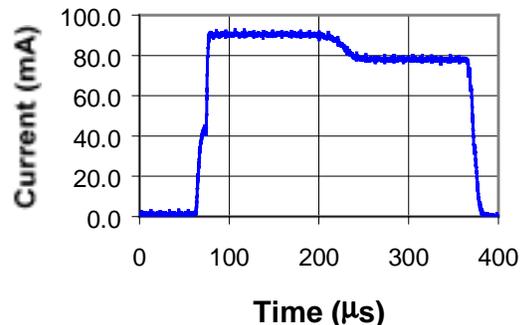

Figure 4: RFQ output beam current vs. time for a 300-µs-long pulse at ~97% of the design RF-field level.

The transmission drops unexpectedly about 150 μs into the pulse. As we raise the RFQ fields the transmission stays high for longer times. With fields above ~105% of design we no longer observe this drop, even for long pulses and CW operation.

We observe higher-than-expected activation near the high-energy end of the RFQ, consistent with high-energy beam loss. If uncorrected, the frequency of the RFQ drops when it accelerates a high average beam current. The lost beam impinges upon and heats the vane tips, causing them to expand inward, reducing the gap. However, the water-cooling system reacts by increasing the temperature of the outer wall to increase the gap, thereby restoring the resonate frequency. Operating the RFQ with fields about 10% above design greatly reduces the magnitude of this beam loss.

The total RF power does not appear to change when the transmission drops. However, when the transmission drops, the RF fields appear to increase slightly in the last meter of the RFQ as though there was less beam loading in that section [4].

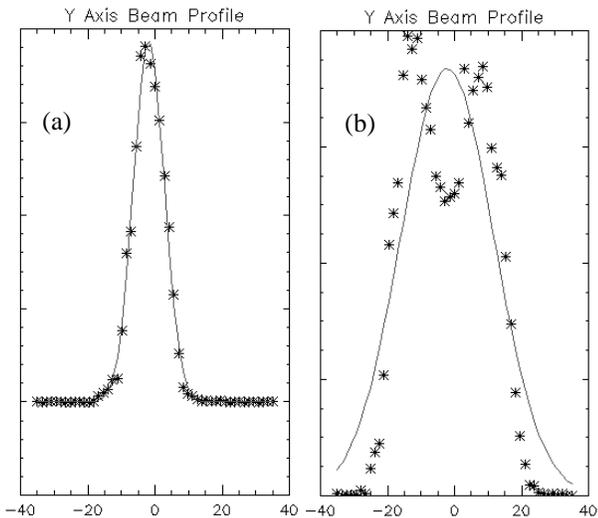

Figure 5: Two vertical wire-scan measurements [19] of beam profile; (a) during the first half, and (b) during last half of beam pulses similar to that shown in figure 4. The HEBT setting was for a Y-emittance scan (near the minimum width in Y) [15] for (a) and (b). The curves are Gaussian fits the data (*).

We theorize that the RFQ fields are trapping low-energy $H^+$ ions near the axis [20]. This extra charge causes the beam size to increase reducing the transmission. This is also consistent with the observation that the beam would cause the beam stop collimator ring to glow whenever the vacuum in the RFQ exceeded about 1-2 $\times 10^{-7}$ Torr. Our conjecture is the beam ionizes more of the residual gas (mostly $H_2$) and the resulting $H^+$ is likely to be trapped in the RFQ bore. Preliminary simulations with a modified version of PARMTEQM, in which an artificial space charge is introduced near the axis, show beam distributions similar to that shown in Figure 5 (b).

At design fields or lower enough beam may strike the vane tips, creating $H^+$ ions that get trapped temporarily in the beam channel. As this trapped charge accumulates the beam becomes larger still until the transmission drops suddenly. Following Ref. [20] to calculate the amount of charge that can be captured both transversely and longitudinally cannot explain the large drop in transmission. However, because the RFQ is 8 m long, a large amount of charge may be captured by the transverse focusing fields temporarily, provided there is a large enough supply of ions. This charge will tend to flow out both ends of the RFQ, but enough charge can accumulate to significantly affect the transmission.

## 5 THE RF POWER AND RESONATE CONTROL SYSTEM

The low-level RF (LLRF) system [21] controls the amplitude, phase, and frequency of the RF power supplied to the RFQ. We used X-ray-endpoint measurements [7,22] to calibrate the fields at 3 different values of the setpoint in the field-control-module (FCM). The resonance control module (RCM) determines the resonant frequency of the RFQ by comparing the phase of the forward power with a sample of the RF in the RFQ. The RCM sends a frequency error signal to the RFQ's water-cooling system. When the resonant frequency error is greater than a specified value (~20 kHz) from 350 MHz, this module switches to a frequency agile mode and synthesizes the frequency required to drive the RF at the RFQ's resonant frequency.

The resonant-control cooling system, [23] using the frequency-error signal from the RCM, controls the temperature of the water flowing to 4 cooling systems on each of the 4 segments. Each of these 4 cooling systems provides water to the outer-wall-cooling channels to maintain the resonant frequency of the RFQ near 350 MHz. Each of these outer-wall-cooling systems has a manually set mixing valve that combines water exiting the outer wall cooling channels with the water provided by the resonant-control system. These manually-set mixing valves provide the compensation for the differential heating of the 4 RFQ segments. A multiplexed system that uses 64 RF pickup probes in the RFQ measures the field amplitude. Four probes in each quadrant of each of the 4 segments provide the field data. After inspecting a plot of the field data the 4 mixing valves are adjusted to make the field distribution at high power nearly the same as the field distribution measured in the RFQ at low power.

## 6 CW OPERATION

On December 17, 1999 we had the first long run with CW beam current of ~100 mA. This run had a few short interruptions but averaged 98.7 mA over 3.3 hr [7]. Key factors that were instrumental in reaching this goal are:

- Reducing the distance between the LEBT solenoid 2 and the RFQ from 30 cm to 15 cm and adding the electron trap at the RFQ entrance.
- Increasing the RF field level in the RFQ to 10% above design reduced the anomalous beam loss in the high-energy end of the RFQ.
- The general improvement in the level of conditioning of the RFQ with operation time. Observations suggest that the vacuum in the RFQ must be about 1.x $10^{-7}$ Torr or lower for good operation.

The RFQ is designed for peak fields at 1.8 Kilpatrick field [24]. When operating at 2 times the Kilpatrick field, the spark rate is not a problem. The estimated spark rate is only 1 to at most 2 sparks per minute average. When a spark is detected, the RF power is turned off for 100 µs. After the RFQ is fully conditioned, most of the beam interruptions are caused by injector arcs, HPRF, or LLRF problems. When these problems and a few others are fixed we see no fundamental reasons why the RFQ can not run for very long periods of time with only short ~100 µs interruptions in the beam.

## 7 SUMMARY

The LEDA RFQ performs as designed, provided the RF field is raised about 10% above the design level to reduce beam loss in the high-energy end of the RFQ and to reach the design transmission. The present RFQ simulation codes do not have the capability of simulating low-energy ions trapped in the RFQ focusing fields. The addition of an electron trap at the entrance to the RFQ is essential to the measurement of the transmission through the RFQ. Simulation of the beam transport through the LEBT with PARMELA allowed understanding the injection of the beam into the RFQ.

The HEBT and beam stop have been moved ~11 m to make room for a 52 quadrupole beam transport line. This beam line will be used to study beam-halo of both matched and unmatched high-current beams [25].

## REFERENCES

[1] D. Schrage *et al.*, "CW RFQ Fabrication and Engineering," Proc. LINAC98 (Chicago, 24-28 August 1998) pp. 679-683.
[2] J. D. Schneider, "Operation of the Low-Energy Demonstration Accelerator: the Proton Injector for APT," Proc. PAC99 (New York, 29 March - 2 April 1999) pp. 503-507.
[3] J. D. Sherman *et al.*, "Status Report on a dc 130-mA, 75-keV Proton Injector," Rev. Sci. Instrum. **69** (1998) 1003-8.
[4] L. J. Rybarcyk *et al.*, "LEDA Beam Operations Milestone and Observed Beam Transmission Characteristics," This conference.
[5] D. E. Rees *et al.*, "Design, Operation, and Test Results of 350 MHz LEDA RF System," Proc. LINAC98 (Chicago, 24-28 August 1998) pp. 564-566.
[6] H. V. Smith, Jr. *et al.*, "Update on the Commissioning of the Low–Energy Demonstration Accelerator (LEDA) Radio–Frequency Quadrupole (RFQ)," Proc. 2nd ICFA Advanced Accelerator Workshop on the Physics of High-Brightness Beams (Los Angeles, CA) (in press).
[7] L. M. Young *et al.*, "Low-Energy Demonstration Accelerator (LEDA) Radio-Frequency Quadrupole (RFQ) Results," *ibid.* (in press).
[8] K. F. Johnson *et al.*, "Commissioning of the Low-Energy Demonstration Accelerator (LEDA) Radio-Frequency Quadrupole (RFQ)," Proc. PAC99 (New York, 29 March - 2 April 1999) pp. 3528-3530.
[9] J. H. Billen *et al.*, "A New RF Structure for Intermediate-Velocity Particles," Proc. 1994 Int. Linac Conf., (Tsukuba, 21-26 Aug. 1994) pp. 341-345.
[10] M. J. Browman and L. M. Young, "Coupled Radio-Frequency Quadrupoles as Compensated Structures," Proc. of the 1990 Linear Accelerator Conference, (Albuquerque, 10-14 Sept. 1990) LA-12004-C, 70.
[11] L. M. Young, "An 8-meter-long Coupled Cavity RFQ Linac," Proc. 1994 Int. Linac Conf., (Tsukuba, 21-26 Aug. 1994) pp. 178-180.
[12] K. R. Crandall *et al.*, "RFQ Design Codes," Los Alamos National Laboratory report LA-UR-96-1836 (Revised August 21, 1998).
[13] L. M. Young, "PARMELA," Los Alamos National Laboratory report LA-UR-96-1835 (Revised January 8, 2000).
[14] L. M. Young, "Simulations of the LEDA LEBT With $H^+$, $H_2^+$, and $E^-$ Particles," Proc. of the 1997 Particle Accelerator Conference (Vancouver, 12-16 May 1997) pp. 2749-2751.
[15] M. E. Schulze *et al.*, "Beam Emittance Measurements of the LEDA RFQ," This conference.
[16] W. P. Lysenko *et al.*, "Determining Phase-Space Properties of the LEDA RFQ Output Beam," This conference.
[17] L. M. Young, "Simulations of the LEDA RFQ 6.7 MeV Accelerator," Proc. of the 1997 Particle Accelerator Conference (Vancouver, 12-16 May 1997) pp. 2752-2753.
[18] J. D. Gilpatrick *et al.*, "Beam Diagnostic Instumentation for the Low-Energy Demonstration Accelerator (LEDA): Comminssiong and Operational Experience," Proc. EPA2000 (Vienna, 26-30 June 2000) (in press).
[19] N. J. Diserens, "Progress in the Development of a 3D Finite Element Computer Program to calculate Space and Image Charge Effects in RF Quadrupoles," IEEE Trans. Nucl. Sci., **NS-32**, (5), 2501 (1985).
[20] M.S. deJong, "Background Ion Trapping in RFQs," Proc. 1984 Linac Conf. (Seeheim, Germany, 7-11 May 1984), pp.88-90.
[21] A. H. Regan *et al.*, "LEDA LLRF Control System Characterization," Proc. LINAC98 (Chicago, 24–28 Aug. 1998) pp. 944-946.
[22] G. O. Bolme *et al.*, "Measurement of RF Accelerator Cavity Field Levels at High Power from X-ray Emissions," Proceedings of the 1990 Linear Accelerator Conference, (Albuquerque, 10-14 Sept. 1990) LA-12004-C, 219.
[23] R. Floersch, "Resonance Control Cooling System for the APT/LEDA RFQ," Proc. LINAC98 (Chicago, 24–28 Aug. 1998) pp. 992-994.
[24] W. D. Kilpatrick, "Criterion for Vacuum Sparking Designed to Include Both rf and dc," Rev. Sci, Instrum., **28**, 824 (1957).
[25] T. P. Wangler, "Beam Halo in Proton Linac Beams," This conference.